\documentclass[aps,prc,superscriptaddress,showpacs]{revtex4}
\usepackage[dvips]{graphicx}
\usepackage{amsmath,amssymb}
\newcommand{\Nt}{\mbox{${\tilde N}_T$}}

\newcommand{\epsi}{\mbox{$\varepsilon$}}

\newcommand{\vx}{\mbox{\boldmath $x$}}

\newcommand{\vz}{\mbox{\boldmath $z$}}

\newcommand{\vp}{\mbox{\boldmath $p$}}

\newcommand{\ls}{\mbox{\LARGE [}}

\newcommand{\rs}{\mbox{\LARGE ]}}

\begin{document}

\title{Monopole Oscillations and Dampings 
in Boson and Fermion Mixture \\
in the Time-Dependent Gross-Pitaevskii and Vlasov Equations}

\author{Tomoyuki Maruyama\footnote{e-mail: tomo@brs.nihon-u.ac.jp}}
\affiliation{
Institute for Nuclear Theory,
University of Washington,
Seattle, Washington 98195, USA}
\affiliation{
College of Bioresource Sciences,
Nihon University,
Fujisawa 252-8510, Japan }
\affiliation{
Advanced Science Research Center,
Japan Atomic Energy Research Institute,Tokai 319-1195, Japan}
\affiliation{Department of Physics, Tokyo Metropolitan University, 
1-1 Minami-Ohsawa, Hachioji, Tokyo 192-0397, Japan}
\author{Hiroyuki Yabu}
\author{Toru Suzuki}
\affiliation{Department of Physics, Tokyo Metropolitan University, 
1-1 Minami-Ohsawa, Hachioji, Tokyo 192-0397, Japan}

\date{\today}

\begin{abstract}
We construct a dynamical model for the time evolution 
of the boson-fermion coexistence system.
The dynamics of bosons and fermions are formulated with
the time-dependent Gross-Pitaevsky equation and the Vlasov
equation.
We thus study the monopole oscillation in the bose-fermi mixture.
We find that large damping exists for fermion oscillations
in the mixed system even at zero temperature.  
\end{abstract}



\maketitle

\maketitle

\section{Introduction}

Over these last several years,
there has been significant progress in the production of 
ultracold gases which realize the Bose-Einstein condensates (BECs) 
\cite{nobel,Dalfovo,becth,Andersen}, 
degenerate atomic Fermi gases \cite{ferG},  
and Bose-Fermi mixtures \cite{BferM}.
These systems offer great promise for studies of new, 
interesting driven-phenomena.

One of the most exciting themes in recent physics is to study
the time-dependent dynamical motion of trapped atoms,
such as collective oscillations \cite{Dalfovo}, 
quantum vortices \cite{vortex} and atomic novae \cite{colap}.
From a theoretical point of view, these phenomena are
very important, as they allow us to construct and examine 
the transport theory in finite many body systems. 
Many body atomic systems are good probes
for such study because the fundamental interactions are 
well understood and can be changed using the Feshbach resonance \cite{Fesh}.

Of the above phenomena collective oscillations are very sensitive to
properties of the system and therefore 
are condensates for studying time-dependent dynamics.
The collective oscillation of BECs have been studied experimentally \cite{ColEx} and 
theoretically \cite{ColBTh}. 
Furthermore, there have been theoretical studies of the Fermi gas in the normal phase \cite{ClFer} 
and in the superconducting phase \cite{ClSp}, 
as well as the Bose-Fermi mixed gases \cite{ClBF,sogo,zeros,Pu}
and the BEC-BCS crossover \cite{BECBCS}.

Collective motions are usually studied with 
the random phase approximation (RPA). 
The RPA however, can only describe minimal vibrations around ground state;
for example it is shown in Ref.\cite{sogo}
that the radial variation is only about 0.05 \%, 
while the amplitudes of actual experimental can be as large as
10 \%.
 
To describe collective oscillations with larger amplitudes
we need to calculate the time evolution of the system
using the time-dependent Gross-Pitaevsky (TDGP)
and  time-dependent  Hartree Fock (TDHF) equations. 
In atomic gas systems, however, the fermion number is greater than 
one thousand, but it is not easy to solve
the TDHF equations with so many fermions and 
limited computer resources.
For system with large fermion number, 
on the other hand,
the semi-classical approximation ($\hbar \rightarrow 0$) becomes useful,
and the Vlasov equation, corresponding to the semi-classical approximation
of the TDHF equations, can be used.

In nuclear physics \cite{BUU1} the Vlasov-Uhling-Uhlenbeck (VUU) approach, 
which is the Vlasov plus two body collision with the Pauli blocking, 
has been introduced, and very nicely explained many kinds of
observables in heavy-ion collisions \cite{BUU2}.  
In addition, this approach has been developed for
the Lorentz covariant framework 
including the non-local mean-fields \cite{TOMO1}.  
As for the boson gas
the time evolution of condensed bosons and thermal bosons have also been
described with the time-dependent Gross-Pitaevsky equation 
and the Boltzmann equation
\cite{JackZar}.
Furthermore C.~Menotti et al.  \cite{Menotti}
studied the expansion of a Fermi gas 
by calculating the Vlasov equation with the scaling method.

In this work we construct a transport model including the condensed bosons 
and fermions based on the TDGP and Vlasov equations.
As a first step we study the collective monopole motion of a the
spherically symmetric system at zero temperature.
This calculation is a good test for the comparison between the RPA and
direct time evolution
calculations.  
In addition one pays attention to the monopole oscillation
from the point of the view of the collapse  so as to know
 compressing processes of systems \cite{colCO}.

In section 2, we describe our transport model. 
In Sec. 3 we give our results for monopole oscillations
in the bose-fermi mixture, and discuss their properties.
We then summarize in Sec. 4.

\section{Formalism}

Here we briefly explain our formalism.
In this work we consider the spherical boson-fermion mixing gas at zero temperature.
 
First we define the Hamiltonian for boson-fermion coexistence systems
as follows.
\begin{equation}
H = H_B + H_F  + H
\end{equation}
with
\begin{eqnarray}
H_B &=& \int d^3 x \left[ - \frac{1}{2} \phi^{\dagger}(\vx) \nabla^2 \phi(\vx)
+ \frac{1}{2} \vx^2 \phi^{\dagger}(\vx) \phi(\vx)
+ \frac{g_B}{2} \{ \phi^{\dagger} (\vx) \phi (\vx) \}^2 \right]  ,
\\
H_F &=&  \int d^3 x \left[
- \frac{1}{2m_f} \psi^{\dagger} (\vx) \nabla^2 \psi (\vx)
+ \frac{1}{2} m_f \omega_f^2 \vx^2 \psi^{\dagger}(\vx) \psi (\vx) \right]  ,
\\
H_{BF} &=& h_{BF}  \int d^3 x \left[
\phi^{\dagger} (\vx) \phi (\vx) \psi^{\dagger} (\vx) \psi (\vx)  \right] ,
\end{eqnarray}
where $\phi$ and $\psi$ are boson and fermion fields, respectively.

The fermion mass $m_f$ and  trapped frequency $\omega_f$ 
are normalized
with the boson mass $M_B$ and the boson trapped frequency $\Omega_B$, respectively.
The coordinates are normalized by $\xi_B = (\hbar / M_B \Omega _B)^{1/2}$.
The coupling constants $g_B$ and $h_{BF}$ are given as
\begin{eqnarray}
g_B &=&  4 \pi a_{BB} ~ \xi_B^{-1},  \\
h_{BF} &=&  2 \pi a_{BF} ~ \xi_B^{-1} (1 + m_f^{-1}),
\end{eqnarray} 
where $a_{BB}$ and  $a_{BF}$
are the scattering lengths between two bosons
and between the boson and the fermion, respectively.

In this work we only consider only zero-temperature, so that
we can approximate the system with only  the condensed bosons
and degenerate fermions.
Namely the total wave function is written as
\begin{equation}
\Phi (\tau) = \{ \prod_{i=1}^{N_B} \phi_c (\vx_i) \} \Psi_f ,
\end{equation}
where  $\tau$ is the time normalized by $\Omega_B^{-1}$, 
$\phi_c$ is a wave function of the condensed boson, and
$\Psi_f$ is a Slater determinant of fermions with single particle wave functions
$\psi_n$.
  
The time evolution of the wave functions are obtained from the variational
condition that
\begin{equation}
\delta \int^{\tau_2}_{\tau_1} d \tau L ({\tau})
= \delta \int^{\tau_2}_{\tau_1} d \tau 
<\Phi(\tau)|\{ i\frac{\partial}{\partial \tau} - H \} |\Phi (\tau) > = 0 .
\label{vari}
\end{equation}
From this condition we derive coupled equations
of the TDGP and TDHF equations as follows.
\begin{eqnarray}
i \frac{\partial}{\partial \tau}  \phi_c (\vx, \tau) &=&
\left\{ - \frac{1}{2} \nabla^2 + U_B (\vx) \right\} ~ \phi_c (\vx, \tau) ,
\label{TDGP}
\\
i \frac{\partial}{\partial \tau}  \psi_n (\vx, \tau) &=&
\left\{ - \frac{1}{2 m_f} \nabla^2 + U_F (\vx) \right\} ~\psi_n (\vx, \tau)
\label{TDHF}
\end{eqnarray} 
with 
\begin{eqnarray}
U_B (\vx) &=& \frac{1}{2} \vx^2 + g_B \rho_B (\vx)
+ h_{BF} ~ \rho_F (\vx) ,
\label{uB}
\\
U_F (\vx) &=& \frac{1}{2} m_f \omega^2_f \vx^2 + h_{BF} ~ \rho_B (\vx) ,
\label{uF}
\end{eqnarray} 
where $\tau$ is the time normalized by $\Omega_B^{-1}$ and 
$\rho_B$ and $\rho_F$ are boson and fermion densities which are given as
\begin{eqnarray}
\rho_B (\vx) &=& N_B |\phi_c (\vx)|^2 ,
\label{rhoB}
\\
\rho_F (\vx) &=& \sum^{occ}_{n} |\psi_n (\vx)|^2 .
\label{rhoF}
\end{eqnarray} 

The number of fermions are too large to solve the above TDHF equations
directly, so instead one uses
the semi-classical approach. 
In the semi-classical limit ($\hbar \rightarrow 0$)
the TDHF equation is equivalent to  
the following Vlasov equation  \cite{KB}:
\begin{equation}
\frac{d}{d \tau} f(\vx,\vp;\tau) =
\left\{ \frac{\partial}{\partial \tau} + \frac{\vp}{m_f}{\nabla_x} -
 [\nabla_x U_F(\vx)][\nabla_p] \right\} f(\vx,\vp;\tau) = 0 ,
\label{Vlasov}
\end{equation}
where $f(\vx,\vp;\tau)$ is the fermion phase-space distribution function 
defined  as
\begin{equation}
f(\vx,\vp,\tau) = \int {d^3 z} <\Phi|
\psi(\vx+\frac{1}{2}{\vz},\tau)
\psi^{\dagger}(\vx-\frac{1}{2}{\vz},\tau) |\Phi>
 e^{-i \vp \vz / \hbar} .
\end{equation}

As an actual numerical method we introduce
the collective coordinate method and the test particle method \cite{TP}
to solve the TDGP equation (\ref{TDGP}) and 
the Vlasov equation (\ref{Vlasov}), respectively.

The wave function of the condensed boson $\phi_c$ is expanded 
with the $s$-wave harmonic oscillator wave function $u_n(\vx)$ as
\begin{eqnarray}
&&\phi_c (\vx,\tau) = \sum_{n=0}^{N_{base}-1} A_n e^{i \theta_n} u_n(b\vx^2) 
e^{-\frac{i}{2} \nu\vx^2} , 
\label{c-tran}
\end{eqnarray}
where $N_{base}$ is the number of the harmonic oscillator bases,
and $b$, $\nu$, $A_n$ and $\theta_n$ are the time-dependent variables.

As for the fermion
the phase-space distribution function is described as
\begin{equation}
f(\vx,\vp,\tau) = \frac{(2 \pi)^3}{\Nt} 
\sum_{i=1}^{{\tilde N}_T N_F} \delta\{\vx-\vx_i(\tau)\} \delta\{\vp-\vp_i(\tau)\} ,
\label{TP-Wig}
\end{equation}
where {\Nt} is number of test-particles per fermion.

Then  we define  the Lagrangian with these coordinates as
\begin{equation}
L(A_n, \theta_n, b, \nu; \vx_i, \vp_i ) = 
 N_B \int d^3 x 
\phi_c^{*} (\vx) i \frac{\partial}{\partial \tau} \phi_c (\vx)
+ \frac{1}{{\tilde N}_T} \sum_{i=1} \{ \vp_i \frac{d}{d \tau} \vx_i \}
- E_T ,
\end{equation}
with the total energy $E_T$  written as
\begin{equation}
E_T = < \Phi | H | \Phi > = E_B + E_F + E_{BF} ,
\label{etot}
\end{equation}
and
\begin{eqnarray}
E_B &=& \int d^3 x \ls \frac{1}{2}
 \phi_c^{*} (\vx) \{ - \nabla^2 + \vx^2 \} \phi_c (\vx)
+ \frac{g_B}{2} \{ \phi_c^{*} (\vx) \phi_c (\vx) \}^2 \rs ,
\label{ebos}
\\
E_F &=& \frac{1}{\Nt} \sum_i 
\ls \frac{1}{2 m_f} \vp_i^2 + \frac{1}{2} m_f \omega_f^2 \vx_i^2 \rs ,
\label{efer}
\\
E_{BF} &=&  \frac{1}{\Nt} h_{BF} \sum_i \rho_B (\vx_i) .
\label{ebf}
\end{eqnarray}
In the above equations
we take $A_n, \theta_n, b, \nu$  and $\vx_i, \vp_i$ 
as time-dependent variables.

The time evolution of these variables are given by 
the Euler-Lagrange equations with respect to 
the time-dependent variables.
For example,
the Euler-Lagrange equations for the fermions
with respect to $\vx_i$ and $\vp_i$, become the following
classical eq. of motion for the test particles,
\begin{eqnarray}
\frac{d}{d \tau} \vx_i (\tau) &=& \frac{\vp_i}{m_f},
\label{eqM1} \\
\frac{d}{d \tau} \vp_i (\tau) &=& - \nabla U_F(\vx).
\label{eqM2}
\end{eqnarray}
We can also obtain the same equations 
by substituting eq.(\ref{TP-Wig}) into eq.(\ref{Vlasov}).
Note here that the test-particle motion describes the
time evolution of the fermion gas, and  does not correspond to
actual single particle motion.

\section{Results}

In this section we show the results of our calculation 
on the monopole oscillation 
in the boson-fermion mixture.
In order to know the characteristic behavior of the boson-fermion 
mixed system,
we consider only the system where the boson number $N_b$ is 
much larger than the fermion number $N_f$ ($N_b \gg N_f$),
which optimizes the overlap of bosons and fermions.
This is because 
the condensed boson density tends to be located in the central region
while the fermion density distributes over a large region due to
Pauli blocking.

In this work we deal with the system $^{39}$K~-$^{40}$K, where
the number of the bosons ($^{39}$K) and the fermions ($^{40}$K)
are $N_b = 100000$ and $N_f = 1000$.
The mass difference between the two atoms is omitted,
$m_f=1$. 
The trapped frequencies are taken as $\Omega_B = 100$ (Hz) and 
$\omega_f = 1$.
The boson-boson interaction parameter  $g_B$ is taken 
to be $g_B = 1.34 \times 10^{-2}$,  which corresponds to 
$a_{BB} = 4.22$(nm) \cite{miyakawa};
in addition we vary the boson-fermion interaction parameter $h_{BF}$
which is measured in units of
$h_0 = 7.82 \times 10^{-3}$ corresponding to 
$a_{BF} = 2.51$(nm) \cite{miyakawa}.
In the numerical calculation we take the number of the harmonic
oscillator bases for the condensed boson to be
$N_{base}=11$ and the number of the test-particle per fermion to be
${\tilde N} = 100$.

\subsection{Ground State}

Before performing the numerical calculation of collective motion, 
we explain the method to construct the ground state 
under the above formalism.
The ground state is defined as the stationary state
with the lowest energy.

In the stationary condition the wave function of the condensed boson
satisfies $\nu = \theta_n = 0$, and 
$\partial E_T/\partial b = \partial E_T/ \partial A_i = 0$.
We can obtain the wave function of the condensed bosons 
by repeating the processes of solving these stationary conditions
and diagonalizing the Hamiltonian matrix with the base of the harmonic
oscillator wave functions.

The fermion phase-space distribution function is given by
the Thomas-Fermi approximation,
\begin{equation}   
f (\vx,\vp ) = \theta[ \mu_f - \epsi(\vx,\vp) ]
\end{equation}
with
\begin{equation}
 \epsi(\vx,\vp) = \frac{1}{2m_f} \vp^2 + U_F (\vx) .
\end{equation}
Here we iterate solving the boson wave function and varying the fermi
energy $\mu_f$ to give the correct fermion number.

For numerical simulations we distribute the test-particles to reproduce
the phase-space-distribution solved in the Thomas-Fermi approximation.
In this process we have to insure the stability of
the ground state.

\begin{figure}[ht]
\hspace*{1cm}
{\includegraphics[scale=0.7]{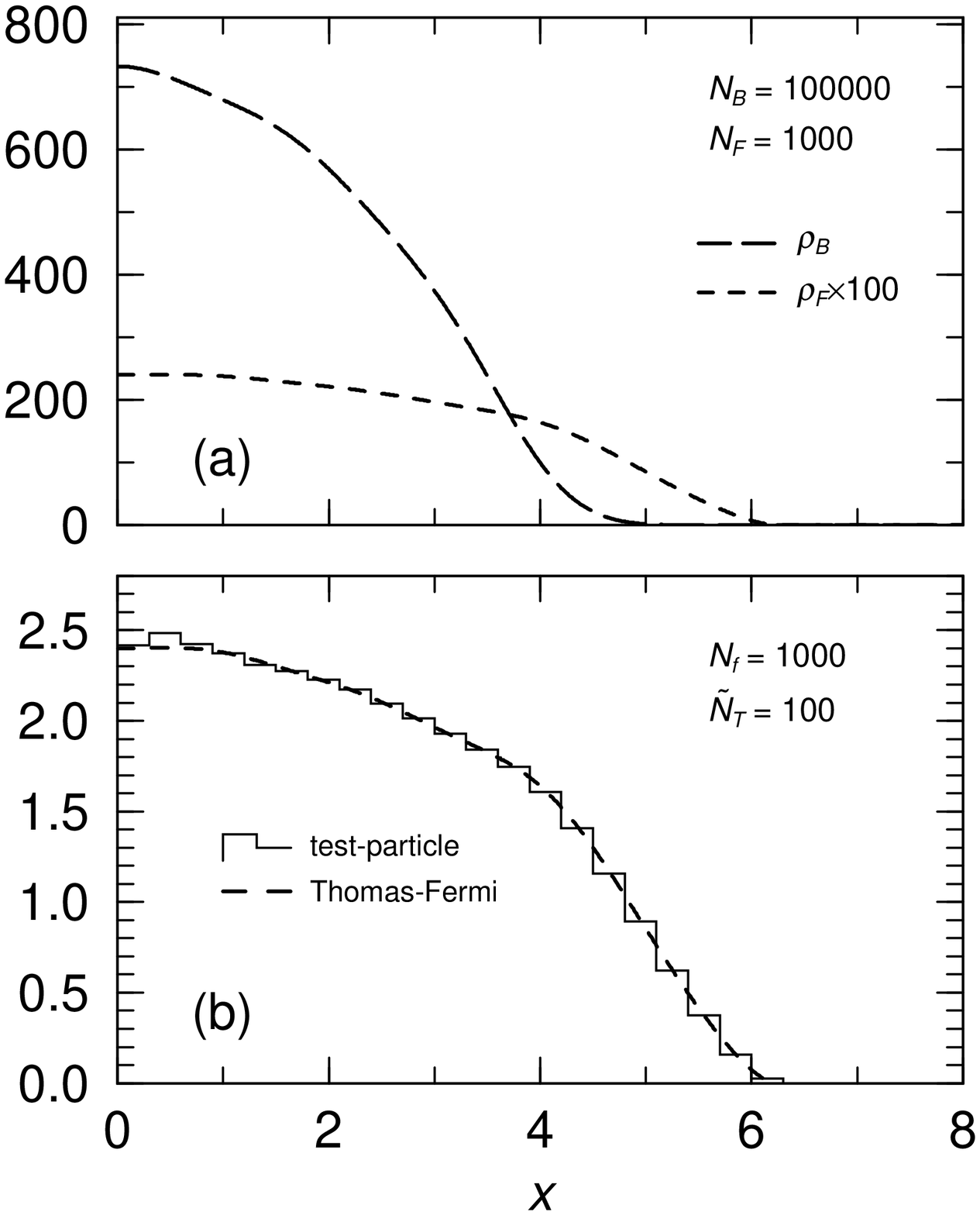}}

\caption
{\small The density-distribution of ground state of the boson ($K^{39}$) and fermion ($K^{40}$)
mixed system with $h_{BF}=h_0$.
The results of $N_B = 10000$ and  $N_F = 1000$ are shown in
Fig.~1a.
The dimensionless distacnce $x$ is in units of $\xi_B$.
The long-dashed and dashed lines indicate the densities
of boson and fermion, respectively.
In lower column the histogram show the results of the test particle.  
}
\label{GRD}
\end{figure}

In Fig.~\ref{GRD}a we show the density distributions of the bosons 
and the fermions (multiplied one hundred)
with the long-dashed and dashed lines, respectively.
We take the boson-fermion coupling to be $h_{BF} = h_0$.
We see a large overlap region between boson and fermion densities.
In addition we show test-particle distribution with ${\tilde N}_T = 100$
in Fig.~\ref{GRD}b, where the histogram indicates the test-particle distribution 
while the dashed line denotes the density in the Thomas-Fermi approximation.
We see that the two results very nicely agree.

\begin{figure}[ht]
\hspace*{0.3cm}
{\includegraphics[scale=0.6,angle=270]{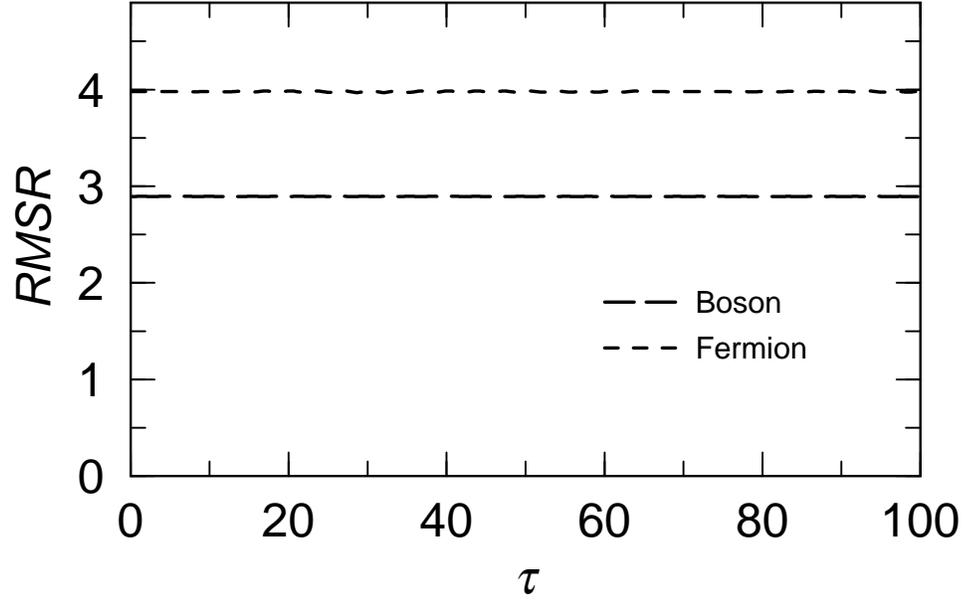}}
\caption{\small
The time-dependence of the root-mean-square radii normarized by
$\xi_B$ 
for $N_B = 100000$ and  $N_F = 1000$ with $h_{BF} = h_0$.
The dimensionless time  $\tau$ is in units of $\Omega_B^{-1}$.
The long-dashed and dashed lines indicate the boson and fermion
root-mean-square radii (RMSR), respectively.
}
\label{GRtev}
\end{figure}

As a next step  we examine the stability of the ground state
by performing a time evolution starting from above the ground state.
In Fig.~\ref{GRtev} we  show the time dependence of 
the root-mean-square-radius (RMSR) for the bosons ($R_b$) 
and the fermions ($R_f$) with long-dashed and dashed lines, respectively.
These results show that their RMSRs
vary only within 0.01 \% over the entire time-evolution, and 
confirm that the ground state is very stable.
Hence we can get very stable and reasonable ground states for the
dynamical calculations.

\subsection{Monopole Oscillations}

In this subsection we show our calculational results for the
monopole oscillations.
We take the boson-fermion coupling 
to be $h_{BF} = h_0$ in all calculations in this subsection.

At the beginning of the numerical simulations 
we scale the boson wave function $\phi_c$ and 
the fermion phase-space distribution function from those in the ground state
as follows.
By using the scaling parameter $s_b$ and $s_f$,
we scale the boson parameter $b$ and the position ($\vx_i$) 
and momentum ($\vp_i$) coordinates of the $i$-th fermions
as 
\begin{eqnarray}
b(0) = b^{(g)} / \sqrt{s_b} , &&
\\
\vx_i(0) = s_f \vx_i^{(g)} , &~~~~~~~&
\vp_i(0) =  \vp_i^{(g)}/s_f ,
\end{eqnarray}
where superscripts $(g)$ represent the coordinates in the ground state.
Then we perform the numerical calculation of the monopole oscillations
with various initial conditions.

In order to see appearance of the monopole oscillation, we define 
\begin{equation}
\Delta x_i (\tau) = R_i (\tau) /{R}^0_i - 1  ~~~~~~ (i = B, F) ,
\end{equation}
where $R_B (\tau)$ and $R_F (\tau)$ are the RMSRs of the bosons and
fermions at time $\tau$, and
$R^0_B$ and $R^0_F$
are the RMSRs in the ground state obtained by the Thomas-Fermi approximation.

\begin{figure}[ht]
{\includegraphics[scale=0.75]{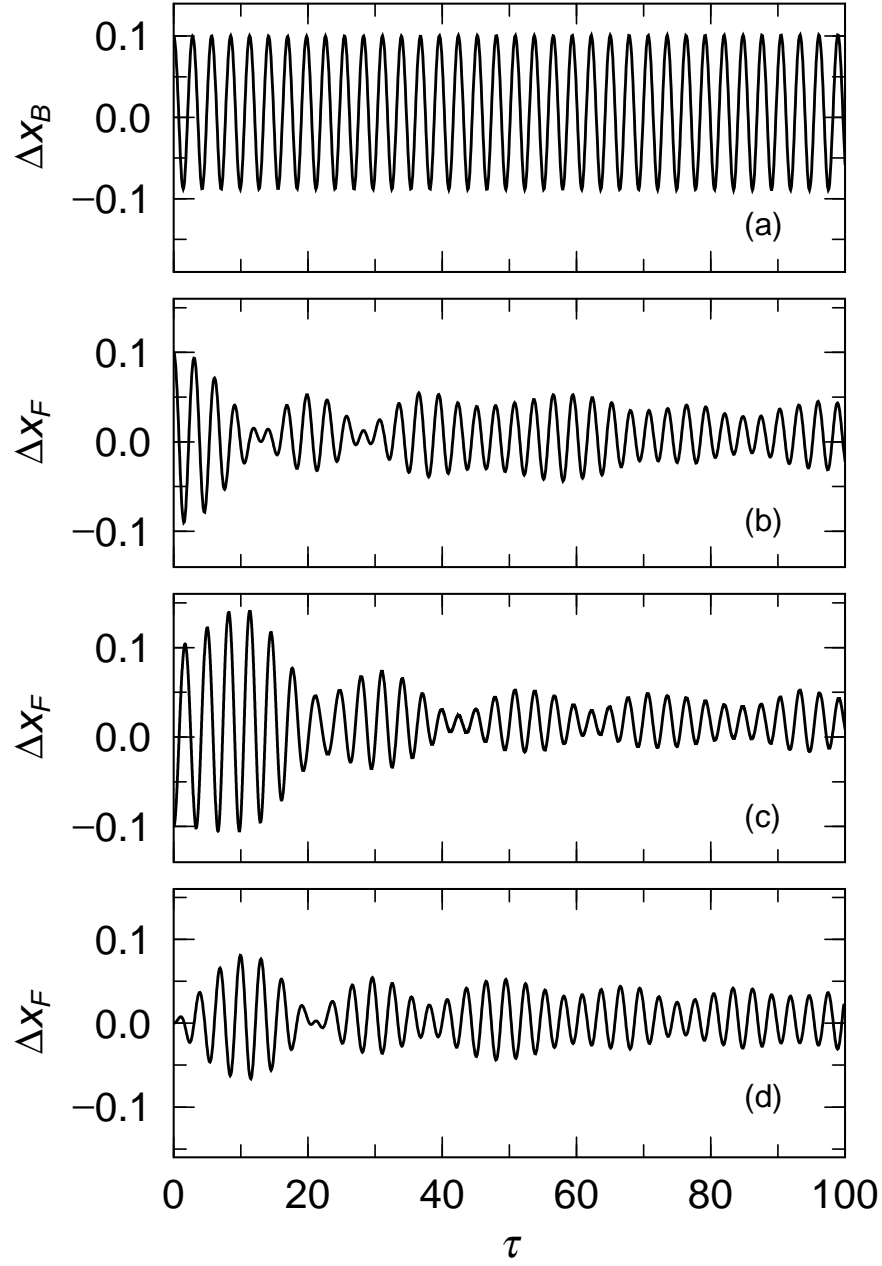}}
\caption{\small
Time evolution of $\Delta x_B$  (a) and $\Delta x_F$  (b)
with $h_{BF} = h_0$ 
at the initial condition that $\Delta x_B(0) = 0.1$ 
and $\Delta x_F(0) = 0.1$.
Time evolution of $\Delta x_F$
at the initial conditions that $s_b = 1.1$
and $\Delta x_F(0) = -0.1$ (c),  and $\Delta x_F(0) = 0$ (d). }
\label{rmTDF2}
\end{figure}

In Fig.~\ref{rmTDF2}  we show the time-dependence of 
$\Delta x_B$ (a) and $\Delta x_F$ (b)
with the initial conditions  $\Delta x_B(0)=0.1$ ($s_b = 1.1$) 
and  $\Delta x_F(0)=0.1$ ($s_f = 1.1$) (in-phase).
In addition we exhibit results with other initial conditions:
$\Delta x_B(0)=0.1$ ($s_b = 1.1$) and $\Delta x_F(0)=-0.1$  
($s_f = 0.9$) (out-of-phase) in Fig.~\ref{rmTDF2}b, 
$\Delta x_B(0)=0.1$ ($s_b = 1.1$) and $\Delta x_F(0)=0$ ($s_f = 1.0$) 
 in Fig.~\ref{rmTDF2}d.
The condensed bosons oscillate independently of the fermions, 
so that the difference between $\Delta x_B$ at these three initial
conditions is not visible.

In the above three cases we see that
the fermion oscillations are forced vibrations with a beat,
which gradually becomes blurred.
This blurring cannot be explained within the RPA.

\begin{figure}
\includegraphics[scale=0.75]{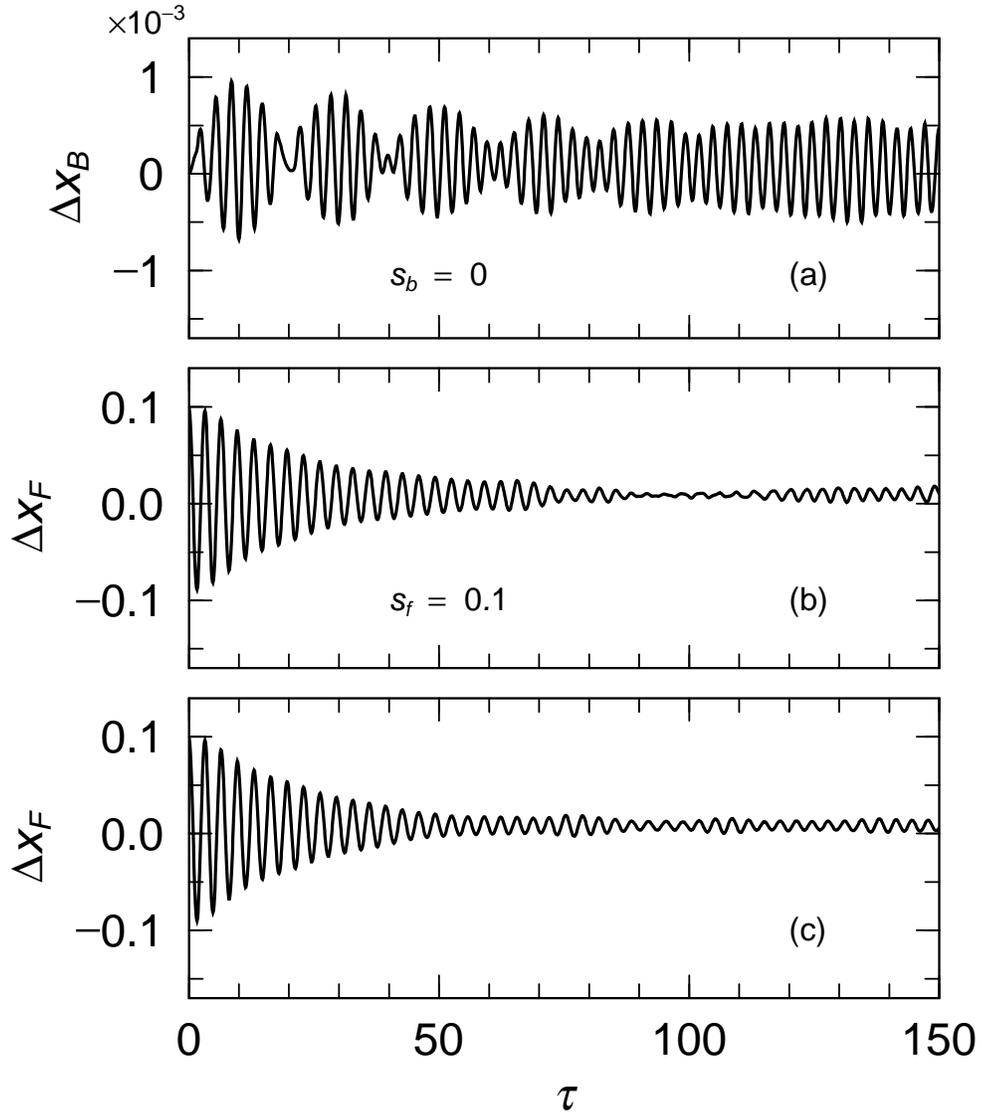}
\caption{\small 
Time evolution of $\Delta x$ for boson (a) 
and fermion (b) with $h_{BF} = h_0$ 
at the initial condition that $\Delta x_B(0) = 0$
and $\Delta x_F(0) = 0.1$.
In the last column
time evolution of $\Delta x$ for fermion (c) 
with freezing the boson motion 
at the same initial condition.}
\label{rmTC1}
\end{figure}

Before discussing this blurring we calculate with another initial 
condition: $\Delta x_B(0)=0$  ($s_b = 1.0$) and
$\Delta x_F(0)=0.1$ ($s_f = 1.1$) .
With this initial condition the boson oscillation is weak and does not
strongly affect the fermion oscillation, see Fig.~\ref{rmTC1}.
The boson oscillation has a beat which appears
 in early time and blurs in latter time stages, 
and the fermion oscillation has a strong damping.
In Fig.~\ref{rmTC1}c 
we  show the results of the calculation 
at the same initial condition, 
but with the boson motion frozen;
namely the fermions move in the fixed  potential $U_F (\vx)$ in the 
ground state.
The damping behavior of the fermion oscillation is almost the same as
that in Fig.~\ref{rmTC1}b,
except that the fermion oscillation is monotonously damped
while the oscillation in Fig.~\ref{rmTC1}b undergoes one beat.

\begin{figure}
\includegraphics[scale=0.6]{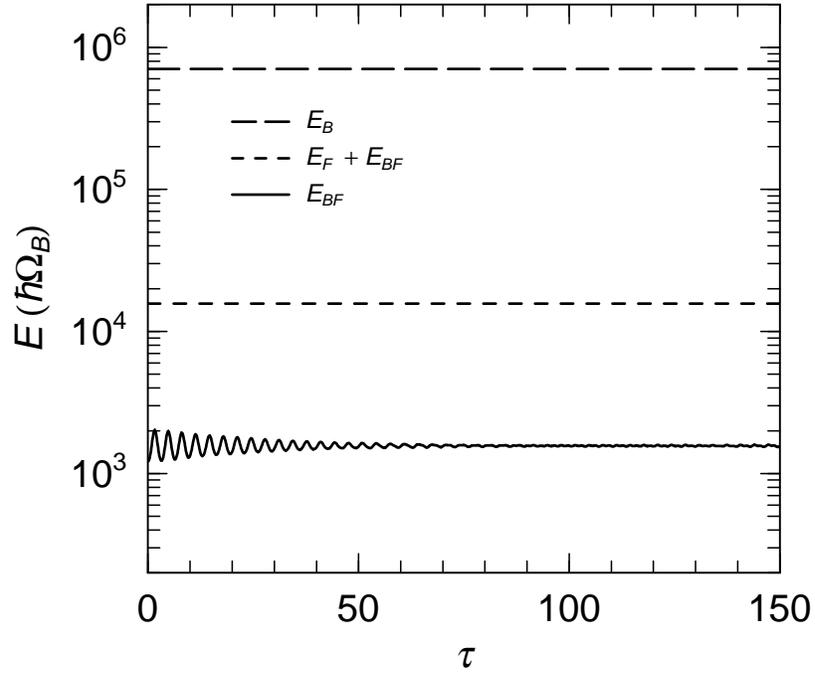}
\caption{\small
Time evolution of the boson energy part $E_B$,
the fermion energy  $E_B$ and  the boson-fermion interaction
energy  $E_{BF}$.
Same as Fig. \ref{rmTC1}}
\label{EngP1}
\end{figure}

This damping process causes
the blurring beat in the fermion oscillation shown 
in Fig.~\ref{rmTDF2}. 
Hence this damping process must play a significant role in collective
motions in the boson-fermion mixtures.
Next we inspect the origin of the damping by focusing on monopole motions
with the initial condition $\Delta x_B(0)=0$.
Here we examine the time-evolution of the energy levels.
In Fig.~\ref{EngP1} we show the time-dependence of $E_B$ (long-dashed), $E_F+E_{BF}$ (dashed) 
and $E_{BF}$ (solid), which are defined in eqs.(\ref{etot}$-$\ref{ebf}).
Note that $E_F + E_{BF}$ corresponds to the sum of the fermion
single particle energy.

First we note that the boson-fermion interaction part, $E_{BF}$,
is much smaller than the boson part, $E_{B}$, and the fermion part, $E_{F}$.
Second the boson part $E_{B}$ and the fermion part, $E_{F}+E_{BF}$, 
do not significantly vary in the time-evolution process.
After the damping ($\tau > 100$)
the energies, normalized by $\hbar \Omega_B$, are given as
\begin{eqnarray}
E_B &=& 7.04 \times 10^{5} ~~~~~ (7.04 \times 10^{5} ), \\
E_F + E_{BF} & = & 1.57 \times 10^{4} ~~~~~ (1.54 \times 10^{4} ), \\
E_{BF} &=& 1.54 - 1.61 \times 10^{3} ~~~~~ (1.57 \times 10^{3}),
\end{eqnarray}
where the values in the brackets indicate the energy contribution to the 
ground state.
The excited energy of the fermions is about 300 ($\hbar \Omega_B$)
while the energy transfer is about 70 ($\hbar \Omega_B$) 
through the boson-fermion interaction.
This fact implies that the damping is not caused by the fermion energy
loss, and in addition that this oscillation state is a many-particle
many-hole state because the excited energy of the one-particle and one-hole
states is about $2 \hbar \Omega_B$ in the monopole oscillations.

Further we calculate the spectrum 
which is obtained by the Fourier transformation
\begin{equation}
F(\omega) = 
\bigg| \int_{T_i}^{T_f} d \tau R^2(\tau) e^{i \omega \tau} \bigg|^2 .
\end{equation}
In order discuss the oscillations,
we vary $T_i$ and $T_f$.

\begin{figure}[ht]
\vspace*{0.5cm}
\includegraphics[scale=0.65,angle=270]{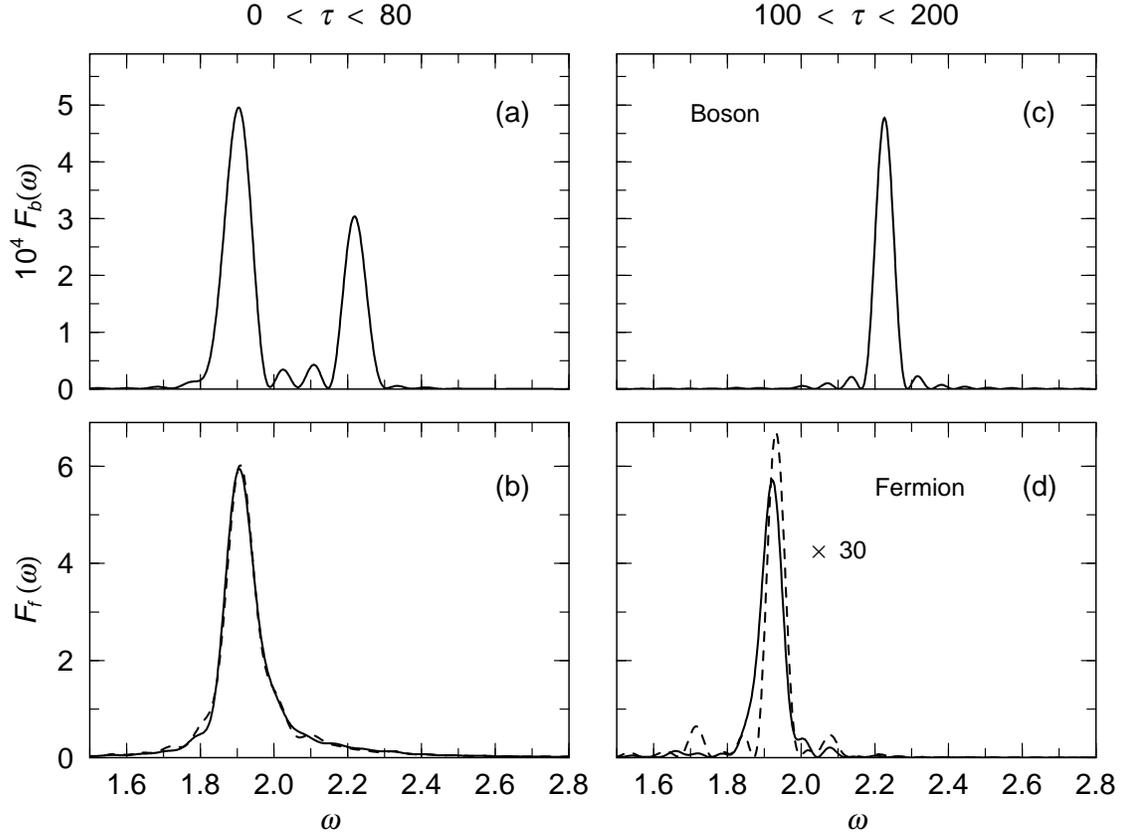}
\caption
{\small 
Spectra of the boson oscillation in upper columns (a,c,e) and
fermion oscillation in down columns (b,d,f).
The spectra deduced from the evolution in $0 < \tau < 30$ 
are shown in the left columns,
that in $30 < \tau < 60$  are in the central columns 
and that in $100 < \tau < 400$ are in the right columns.
The frequency $\omega$ is in unit of $\Omega_B$.
See Fig.~\ref{rmTDF2} for the details of the oscillation.
}
\label{spectA}
\end{figure}

In Fig.~\ref{spectA} we show the results
of the spectra obtained by integrating over
the regions $0 < \tau < 80$ (damping period) for the bosons (a) and
fermions (b), 
and over $100 < \tau < 200$ (after damping ) for the bosons (c) and fermions (d).
In addition we plot the spectra for fermion oscillation
with the boson motion frozen, with the dashed lines in Fig.~\ref{spectA}b and \ref{spectA}d. 

In Fig.~\ref{spectA}b we see that 
the peak position for the fermion spectrum is $\omega \approx 1.91$,
and does not vary in any of the time regions while the width becomes 
slightly narrower after the damping is finished (Fig.~\ref{spectA}d).

On the other hand there are two peaks for the boson spectra;
at the frequency  $\omega \approx 1.91$ and $2.23$ in the fermion damping
process.
After the damping, there is only one boson oscillation mode at $\omega \approx 2.23$,
similar to the fermion oscillation.

We can consider 
the intrinsic frequencies of the boson and fermion oscillations to be
$\omega_M^b = 2.23$ and   $\omega_M^f = 1.91$ in our calculation.
The boson intrinsic frequency agrees with $\omega = \sqrt{5}$ which has been 
shown in the system only including the condensed bosons  \cite{Dalfovo}.
The boson intrinsic mode is almost independent of fermions
because the boson number is much larger than the fermion number.
In addition the frequency of the fermion oscillation 
is the same with or without the boson motion.

In the short time period
both the boson and fermion oscillations can be approximately 
described within the linear response theory as
\begin{eqnarray} 
\Delta x_B & \approx & A_{bb} \cos (\omega_M^b \tau) - 
             A_{bf} \cos (\omega_M^f \tau) , \\
\Delta x_F & \approx & A_{fb} \cos (\omega_M^b \tau) - 
             A_{ff} \cos (\omega_M^f \tau) .
\end{eqnarray}
If the amplitude of the boson oscillation in the initial condition
is not small, 
the boson oscillation negligibly involves a mode with
the fermion intrinsic frequency,
$|A_{bb}| \gg |A_{bf}|$, and the fermion oscillation
becomes the usual forced oscillation $|A_{fb}| \approx |A_{ff}|$
as shown in Fig.~\ref{rmTDF2}. 
 
When $\Delta x_B(0) =0$ such as in Fig.~\ref{rmTC1},
the fermion oscillation does not have a mode with the boson
frequency ($|A_{fb}| \approx 0$).
In the beginnings these fermion oscillations trigger 
the boson oscillations, 
and then the boson oscillations dominantly hold 
the mode with the fermion frequency,  $A_{bb} \gg A_{bf}$.
As the time evolves, the amplitude of the boson oscillation
becomes larger, but the amplitude of the fermion oscillation 
contrarily decreases, and
the $A_{bf}$ is gradually reduced by the damping of
the fermion oscillation.
The amplitude of the fermion oscillation 
becomes too small to influence the boson oscillation,
the boson oscillation loses the mode with the fermion frequency, 
and its beat finally disappears.

The behavior of the fermion oscillation in the full calculation 
is the same as that when the boson motion is frozen, and the fermions
do not lose energy through oscillation.  
Hence the fermions move almost independently each of other, 
and the damping of the fermion oscillations must come from
properties of the fermion single particle motion.  

\begin{figure}
\includegraphics[scale=0.6,angle=270]{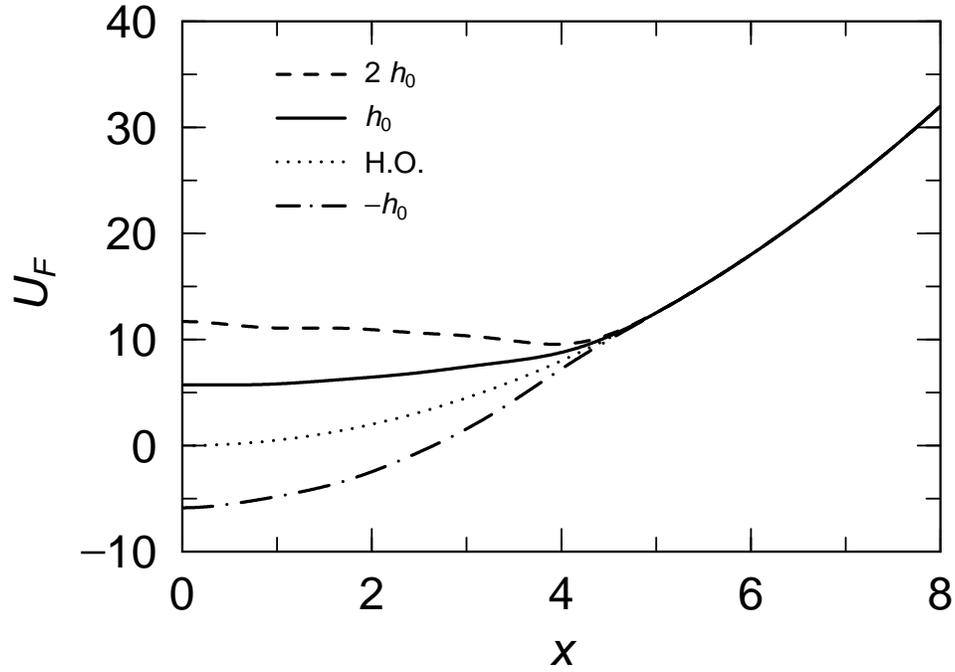}
\caption{\small
Fermion potential at the ground state.
The dashed, solid, and chain-dotted lines show those
at $h_{BF} = 2 h_0$,  $h_{BF} = h_0$ and  $h_{BF} = -h_0$, respectively.
The dotted line indicates the harmonic oscillator potential.}
\label{grdPOT}
\end{figure}

The boson density is distributed in a smaller region than the fermion density,
so the fermion potential does not have simple harmonic oscillator shape,
which is shown with the solid line in Fig.~\ref{grdPOT}.
The Thomas-Fermi approximation, which is available
in the case of the large boson number,
gives the boson density as $\rho_B \approx (2 \mu_B - x^2)/2g_B$, 
where $\mu_B$ is the boson chemical potential.
Then the fermion potential $U_F$ is 
approximately written as
\begin{equation}
U_F(\vx) \approx \left\{ \begin{array}{cc}
                   \frac{h_{BF}}{g_B} \mu_B 
+ \frac{1}{2} (m_f \omega_f^2 - \frac{h_{BF}}{g_B}) \vx^2
& ~~~~(|\vx| < \sqrt{2 \mu_F})~~~~~~ \\
 \frac{1}{2} m_f \omega_f^2  \vx^2 & ~~~~(|\vx| > \sqrt{2 \mu_F})~~~~~
\end{array} \right.  .
\label{UFpot}
\end{equation}
Thus the fermion potential is separated into two region, 
the inside and outside of the boson populated region.

In the single harmonic oscillator potential the fermion motion 
is always harmonic in both the semi-classical and quantum approaches,
but the above fermion potential, which is separated into 
the two regions,
leads to anharmonicity when fermions move in the both regions.

In the quantum calculation this damping is explained with 
the phase factor of the fermion single particle wave functions
given by their single particle energies. 
If the boson motion is frozen, the wave-function of the $n$-th fermion 
is written as
\begin{equation}
\psi_n(\vx,\tau) = \sum_i a_i^{(n)} v_i(\vx) e^{-i \epsi_i \tau},
\end{equation}
where $v_i$ is the wave-function of the $i$-th orbit of the ground state,
$\epsi_i$ is its single particle energy, and $a_i^{(n)}$ is a coefficient.
The expectation value of $\vx^2$ is
\begin{equation}
<\vx^2> = \sum_n \int d^3 x \psi_n^{*} (\vx,\tau) \vx^2 
\psi_n (\vx,\tau) = \sum_{i,j}  C_{i,j}
e^{i( \epsi_j - \epsi_i ) \tau}
\end{equation}
with
\begin{equation}
C_{i,j} = \sum_{n}  a_i^{(n)}  a_j^{(n)*} 
\int d^3 x v_j^{*} (\vx) \vx^2 v_i(\vx) .
\end{equation}
This coefficient is usually negligible when $|i-j| \ge 2$, and then
\begin{equation}
<\vx^2> \approx \sum_i ~\left\{ C_{i,i} +  
C_{i,i+1} e^{i( \epsi_{i+1} - \epsi_{i} ) \tau}.
+C_{i+1,i} e^{-i( \epsi_{i+1} - \epsi_{i} ) \tau} \right\} .
\end{equation}
When the fermion potential is only the harmonic oscillator potential, 
$\epsi_{i+1} - \epsi_{i} = 2\omega_f$, and 
the oscillation of $<\vx^2>$ does not exhibit a damping.    
As shown before, however, the fermion potential includes two kinds
of harmonic oscillator potentials with the different frequency.
In addition the potential in the boson region
is not exactly that of the harmonic oscillator.
Then the difference of the single particle energies, 
$\epsi_{i+1} - \epsi_{i}$, has dependence on $i$,
and the fermion monopole oscillation  
includes various vibration modes with different freqencies, 
and its oscillation shows a damping.
It was shown in Ref.~\cite{Poetting} that 
an oscillation of  population difference in two-component fermi gase has
a  damping because of a similar multimode dephasing.

In the semi-classical calculation, the frequencies of modes including
the oscillation is continuous, though they are discrete 
in the full quantum calculation.
In the system with large particle number, however, 
the semi-classical calculation well describes 
the time evolution of the phase space distribution.
In the harmonic oscillator potential, especially, this time evolution
completely agrees with that in the quantum calculation
because higher order terms with respect to $\hbar$ \cite{KB},
which are omitted in eq.(\ref{Vlasov}), exactly vanish.
Hence our calculations sufficiently describe the actual
time-evolution in the mean-field level.

In numerical simulations such oscillation properties are
described by motions of the test-particles.
A test particle passes over 
both regions, inside and outside of the boson region.
Here we consider oscillation behavior of radial distance
of this $i$-th test particles as $|\vx_i|$.
In a non-harmonic oscillator potential the period between successive
maximum $|\vx_i|$ depends on the orbit.
This anharmonicity gradually reduces the amplitude of $\Delta x$.

As mentioned before, the amplitude is about $\Delta x \sim 10^{-4}$ in
the RPA
calculation and the excited energy is about $2 \hbar \Omega_B$.
In the present calculation, however,
the excited energy when $\Delta x_F(0)=0.1$ ($s_f=1.1$) 
is about $300\hbar \Omega_B$.   
Oscillation states such as those in the present calculation are 
multi-particle and multi-hole states.
Under an anharmonic potential, all test-particles do not oscillate 
with the same period, and hence
this anharmonicity is the cause of the observed damping of $\Delta x_F$.

\subsection{Amplitude and Boson-Fermion Coupling Dependences}

In this subsection we investigate the oscillation behavior 
by changing the initial amplitude of fermion oscillation,
$\Delta x_F(0)$, and the boson-fermion couplings, $h_{BF}$.

\begin{figure}
{\includegraphics[scale=0.7,angle=270]{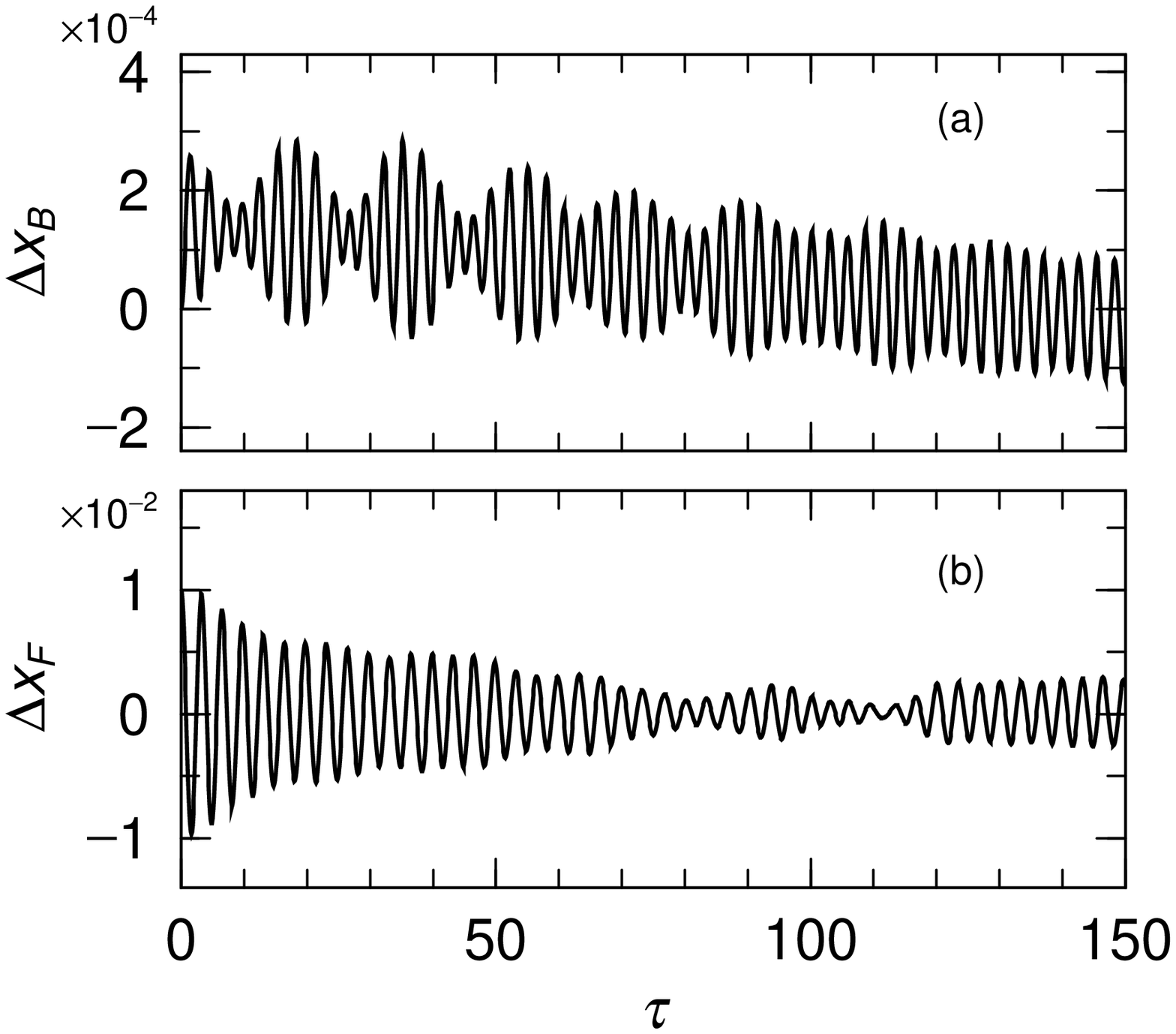}}
\caption{\small
Time evolution of $\Delta x$ for boson (a) 
and fermion (b) 
with $h_{BF} = h_0$
at the initial condition that $\Delta x_B(0) = 0$
and $\Delta x_F(0) = 0.01$.}
\label{rmAMP1}
\end{figure}

First we examine the amplitude dependence of the damping.
In Fig.~\ref{rmAMP1} we show the results with the
initial conditions that $\Delta x_B(0) = 0$ and 
$\Delta x_F(0) = 0.01$ (small amplitude).
The excited energy of this oscillation is about 30 ($\hbar \Omega_B$),
which is still much larger than the one-particle one-hole excited energy.
The strong damping also appears there even with 
such small amplitude oscillation, which becomes about $10^{-3}$ after the damping.
The test particle method cannot however give sufficient accuracy
to exactly describe oscillations with such small amplitudes
after the damping.
Therefore, the more detailed study lies beyond 
the scope of this work.

\begin{figure}
{\includegraphics[scale=0.7,angle=270]{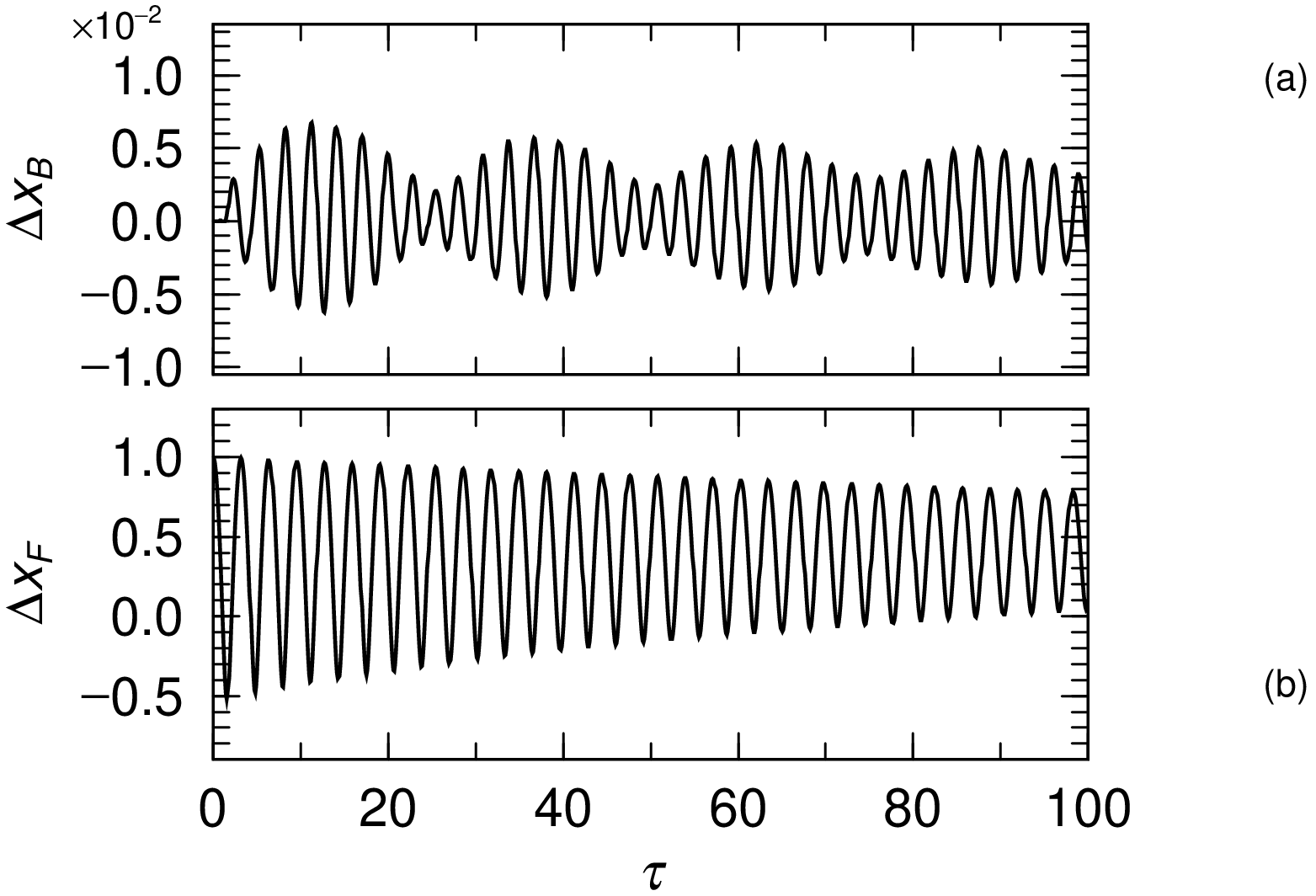}}
\caption{\small
Time evolution of $\Delta x$ for boson (a) 
and fermion (b) with $h_{BF} = h_0$
at the initial condition that $\Delta x_B(0) = 0$
and $\Delta x_F(0) = 1.0$. }
\label{rmAMP2}
\end{figure}

\begin{figure}[ht]
\vspace*{0.2cm}
\hspace*{1.0cm}
\includegraphics[scale=0.7]{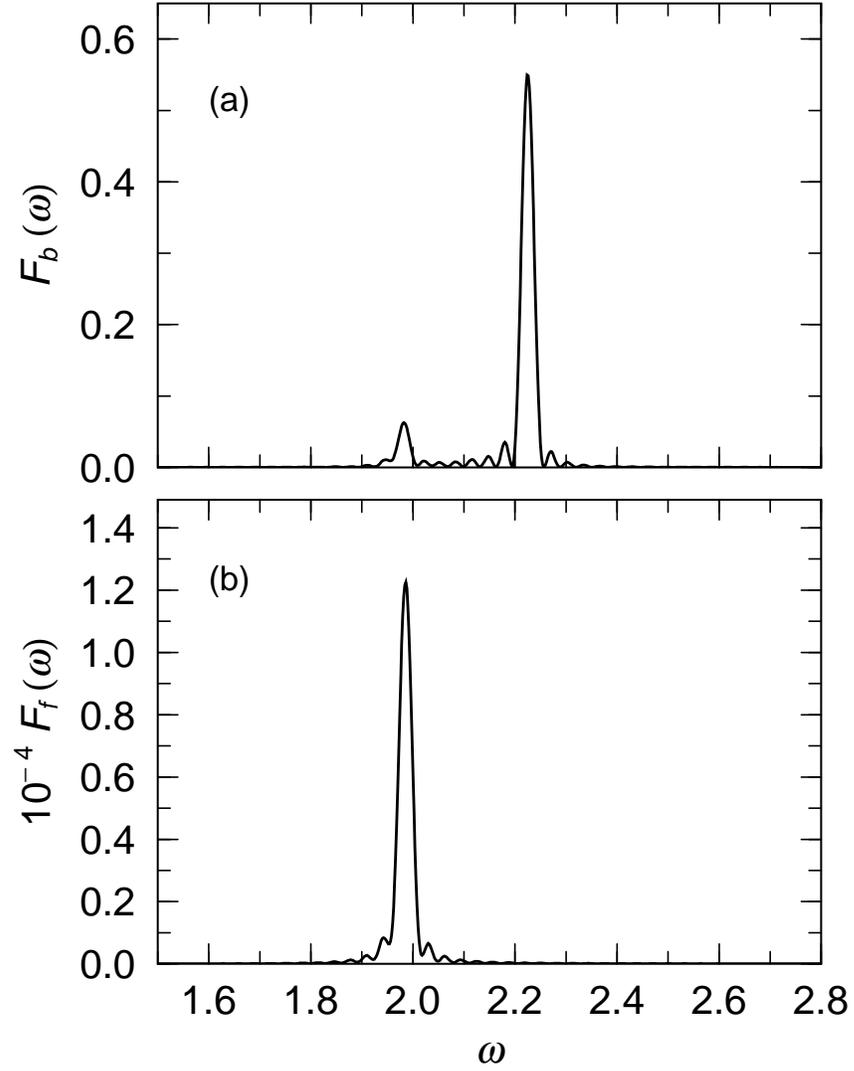}

\caption
{\small 
Spectra of the boson oscillation in upper columns (a) and
fermion oscillation (b) at the initial condition that $\Delta x_B(0) = 0$
and $\Delta x_F(0) = 1.0$.
The spectra are deduced from the evolution in $0 < \tau < 200$ 
in the oscillation shown in Fig.~\ref{rmAMP2}.
}
\label{frAMP2}
\end{figure}

In Fig.~\ref{rmAMP2} we show the results with the
initial conditions that $\Delta x_B(0) = 0$ and $\Delta x_F(0) = 1.0$ 
(large amplitude);
we do not see such strong damping 
though the excitation energy is about 1800 ($\hbar \Omega_B$).
In Fig.~\ref{frAMP2} we show the spectra of the above boson and fermion oscillations
for the large amplitude.
The boson and fermion spectra have peaks at quite different frequencies.
The boson spectrum has a large peak at the boson intrinsic frequency \
$\omega \approx \omega_M^f$ (a),
while the fermion one has a peak at $\omega  \approx 2$ (b),
which is the frequency of the monopole oscillation with only one
harmonic oscillator potential.
In the large amplitude oscillation
most of the fermions have large kinetic energy in the central region
and pass through this region quickly,
so that the fermions mostly stay outside the boson density region.
In large amplitude oscillations,
therefore, the oscillation frequency agree with
that only in the trapped potential and, 
the rapid damping does not appear.
The final amplitude after the damping must be much larger
than that when $\Delta x_F(0)=0.1$. 

\bigskip

\begin{figure}
{\includegraphics[scale=0.7,angle=270]{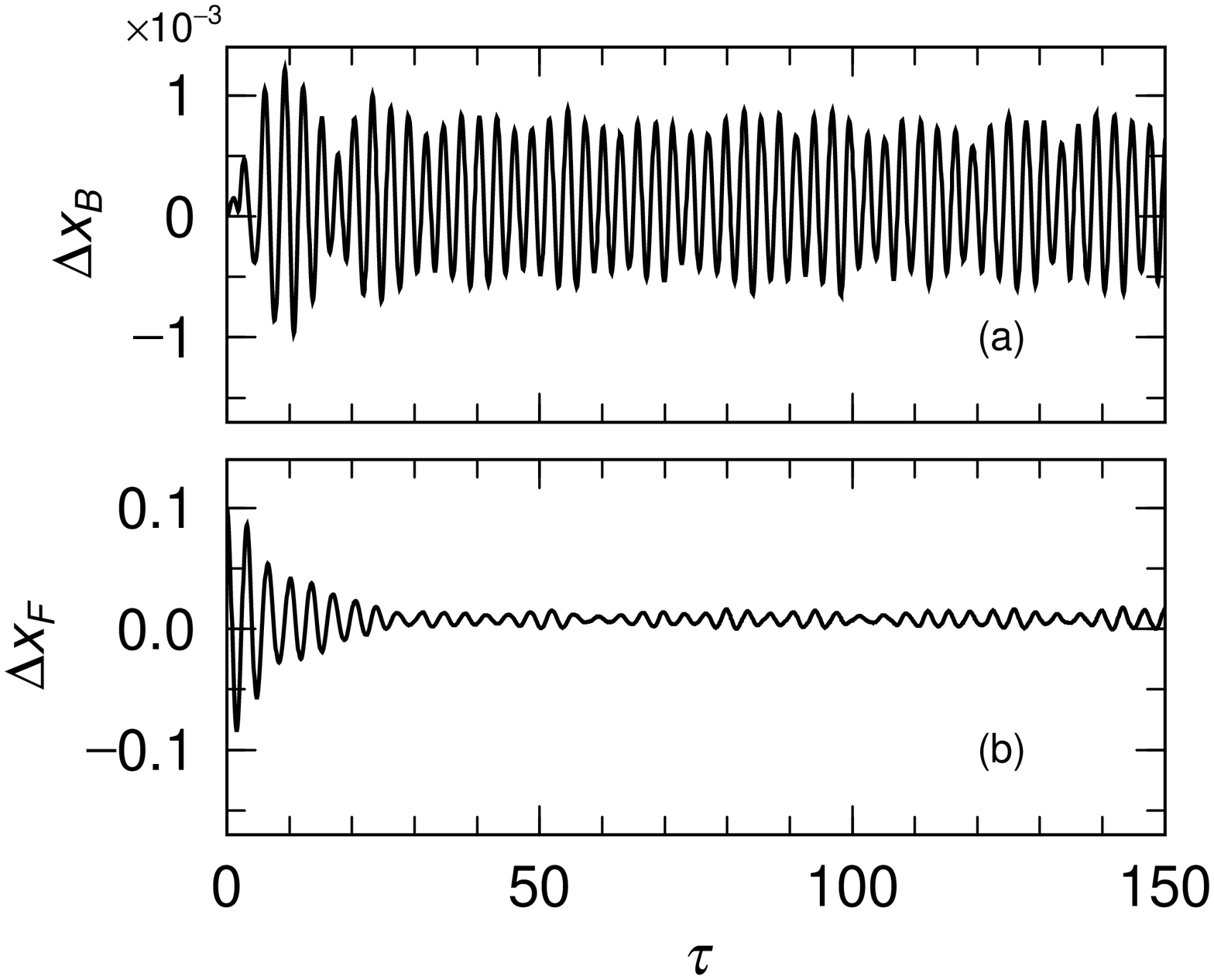}}
\caption{\small 
Time evolution of $\Delta x$ for boson (a) 
and fermion (b) at the initial condition that $\Delta x_B(0) = 0$
and $\Delta x_F(0) = 0.1$ with the boson-fermion coupling $h_{BF} = 2 h_0$ }
\label{rmPP2}
\end{figure}

\begin{figure}
{\includegraphics[scale=0.7,angle=270]{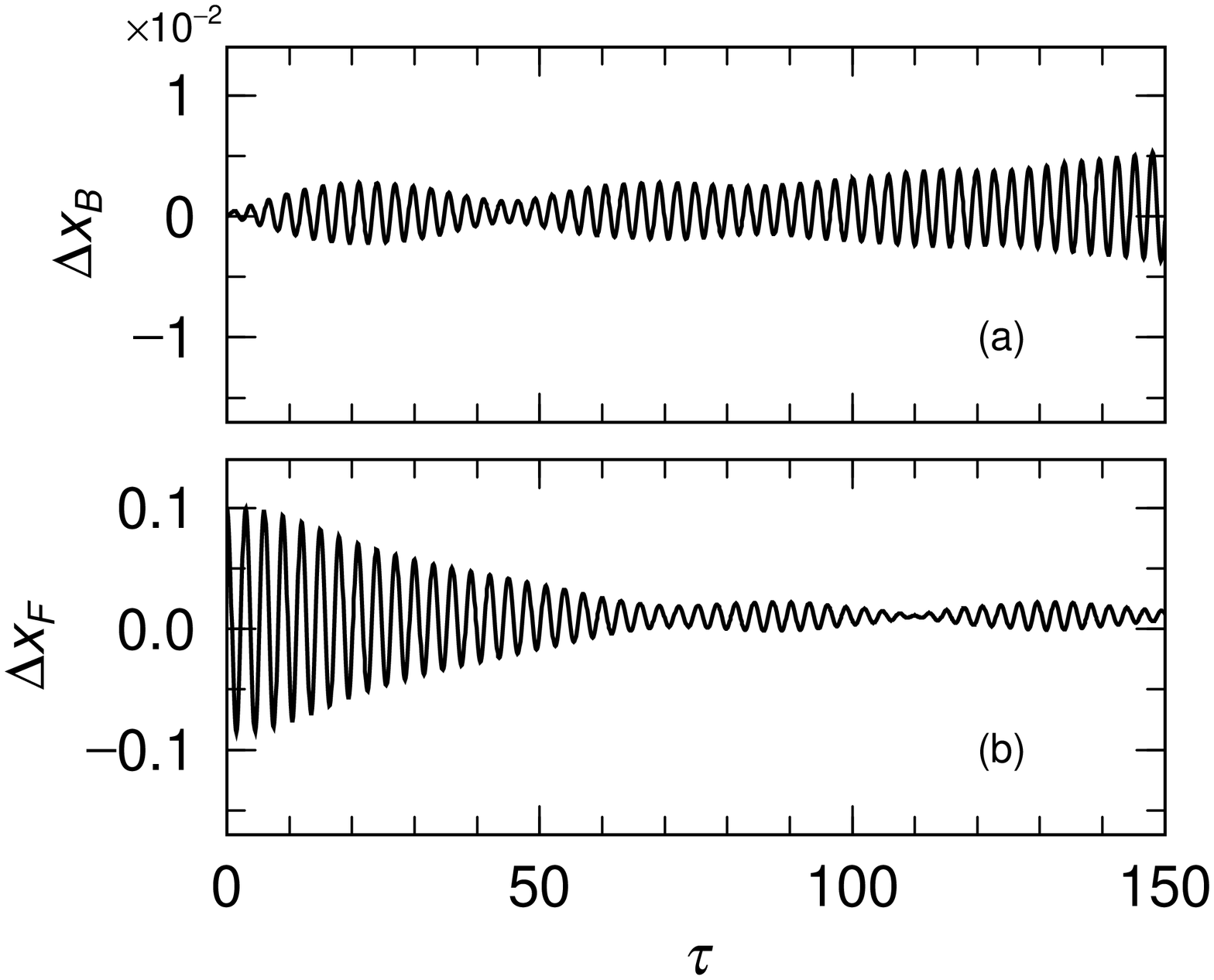}}
\caption
{\small 
Time evolution of $\Delta x$ for boson (a) 
and fermion (b) at the initial condition that $\Delta x_B(0) = 0$
and $\Delta x_F(0) = 0.1$ with the boson-fermion coupling 
$h_{BF} = - h_0$ }
\label{rmPM1}
\end{figure}

Next we vary the strength of the boson-fermion 
coupling $h_{BF}$, and perform calculations of the oscillation
at the same initial condition
$\Delta x_B(0)=0$ and $\Delta x_F(0)=0.1$ . 
We show the results  with $h_{BF}/h_0 = 2$ in Fig.~\ref{rmPP2} and
$h_{BF}/h_0 = -1$ in Fig.~\ref{rmPM1}. 

In Fig.~\ref{rmPP2} the features of the oscillation which have been seen
in Fig.~\ref{rmTC1} are more clear;
the damping of the fermion oscillation 
and the blur of the beat in the boson oscillation 
occurs sooner with $h_{BF} = 2 h_0$ than $h_{BF} = h_0$.

If the boson-fermion coupling is negative, the oscillation behavior is 
a little different.
When $h_{BF} = - h_0$ in Fig.~\ref{rmPM1}, the fermion oscillation 
undergoes beating even after the motion is damped.
Additionally the amplitude of the boson oscillation is much larger than 
that of $h_{BF} > 0$.
This beat of the fermion oscillation after the damping is caused by
the rather large amplitude of the boson oscillation.

As the repulsive interaction between bosons and fermions is increased,
the damping of the fermion oscillation is faster, but 
the amplitude of the boson oscillations do not increase.
In the attractive interaction, however, the damping is still rapid,
but the amplitude of the boson oscillation is large.
We believe this difference  comes from the overlap region between
boson and fermion densities.

When the fermions feel a repulsive force in the boson density region
(see Fig.~\ref{grdPOT}),
the fermions stay outside of this boson region and
have little influence on the boson oscillation.
Then the amplitude of the boson oscillation does not become larger, 
and its beat becomes blurred soon in spite of the stronger interaction.
When the interaction is attractive, on the other hand,    
many fermions moves inside of the boson density region,
and then boson and fermion motions are more significantly coupled.   

\section{Summary}

In this paper we study the collective monopole oscillation of bose-fermi
mixtures, where the number of bosons is much larger than that
of fermions, by solving directly the time-dependent equations. 
When the initial amplitude is about 10 \% of the ground state
RMSR, we find a rapid damping of the fermion oscillation
and a beat in the bosonic vibration at the early time stage while
this damping and beat almost disappear in the later time stage.  
This rapid damping does not exist at zero temperature in the system including
only one kind of boson  \cite{chevy};
it is a typical feature in the bose-fermi mixtures.

Our analysis shows this damping is caused by the anharmonicity of the multi-particle and
multi-hole states.
Since the fermion mean-field is separated into two parts, 
inside and outside of the boson density region,
fermions feel very different forces in these two regions 
and their motions are not harmonic.
Then these motions becomes chaotic, and the oscillation amplitude decreases.

In this work we focus on the oscillation with the initial condition
$\Delta x_B(0) = 0$, 
where only the fermions have initial motion at the beginning.
In this case we can clearly see damping properties of 
the fermion oscillation because the boson oscillation has very small
amplitude.
Even if the amplitude of the initial boson oscillation is larger, the damping also plays
a significant role.
When the fermion oscillations have a large amplitude, it always tends to cause
damping, but the amplitude after the damping is enhanced, again, by the boson
oscillation.
The fermion oscillation repeats the damping and enhancement, and its beat gradually 
blurs; the oscillation behavior is not recursive.

This damping process cannot be described with  RPA, 
where the motion is assumed to be harmonic.
As shown in Ref.\cite{sogo} the excitation energy obtained in RPA
corresponds to  oscillations with  the amplitude of $10^{-4}$ of  RMSR. 
In actual experiments, however, the amplitude is not so small, 
the oscillation states are multi-particle and multi-hole states,
and hence we must solve the time-dependent process directly.

Even if the boson-fermion interactions are attractive, 
both the boson and fermion oscillations have an oscillation behavior
similar to those in the repulsive interaction in the damping time stage.
In the attractive interaction, however, the amplitude of the boson oscillation
slightly increases and the fermion oscillation has a small beat after 
the damping time stage.

In this work we do not show calculations with stronger attractive force
between  bosons and fermions.
If the attractive force is stronger, 
the frequency of the fermion oscillation becomes larger, and 
finally agrees with that of the boson intrinsic oscillation;
where the two oscillation must make a resonance.
We have investigated this phenomena, and found resonances
which have quite different behaviors \cite{tomo2}.

Furthermore we do not take into account two body collisions and
thermal bosons.
Since two fermions do not directly interact, 
dissipation of fermions occurs through collisions between
fermions and thermal bosons.
In the system $N_b \gg N_f$ at zero temperature, however,
the thermal bosons are very few, and then the damping process
shown in this work must be dominant in the early time stage.
After this damping the collisional damping may play a role
in the time-evolution process.
In future we would like to introduce this collisional process
in our simulation.  
   
\bigskip
{\bf Acknowledgement}

We thank Andr{\'e} Walker-Laud for helpful comments on the manuscript.
T.M. thanks the Institute for Nuclear Theory at  University of Washington for the hospitality
and partially support during the completion of this work.

\newpage

\newpage

\bigskip

\hspace{0.5cm}

\newpage

\newpage

\newpage

\newpage

\newpage

\newpage

\newpage

\bigskip

\end{document}